\documentclass[]{aipproc}

\layoutstyle{6x9}
\usepackage{bm}

\def\gsim{\lower.4ex\hbox{$\;\buildrel >\over{\scriptstyle\sim}\;$}} 
\def\lsim{\lower.4ex\hbox{$\;\buildrel <\over{\scriptstyle\sim}\;$}}

\def\Om{{ \Omega}}
\def\beg{\begin{eqnarray}}
\def\ende{\end{eqnarray}}
\def\lsim{\lower.4ex\hbox{$\;\buildrel <\over{\scriptstyle\sim}\;$}}

\usepackage{graphicx}

\begin{document}

\title{Taylor-Couette flow: MRI, SHI  and SRI\footnote{dedicated to Prof. E.P. Velikhov on the occasion of his  70th birthday}}

\author{G\"unther R\"udiger}{
  address={Astrophysikalisches Institut Potsdam, An der Sternwarte 16, D-14482 Potsdam, Germany}
}

\begin{abstract}
The linear stability theory of Taylor-Couette flows (unbounded in $z$) is described including  magnetic fields, Hall effect or  a density stratification  in order to prepare laboratory experiments to probe the stability of differential rotation in astrophysics. If an axial field is present then the magnetorotational instability (MRI) is investigated  also for small magnetic Prandtl numbers.  For rotating outer cylinder beyond the Rayleigh line characteristic minima are found for magnetic  Reynolds number of the order of 10 and  for Lundquist  numbers of order 1. The inclusion of extra toroidal current-free fields leads to new  oscillating solutions with rather low Reynolds numbers and Hartmann numbers. The Hall effect establishes an unexpected  `shear-Hall instability' (SHI) where shear and magnetic field have opposite signs. In this case even rotation laws increasing outwards may become unstable. Recently global solutions  have been found for the Taylor-Couette flows with density stratified in axial directions (`stratorotational instability', SRI). They exist beyond the Rayleigh line for Froude numbers of moderate order but also not for too flat rotation laws with $\hat\eta^2 < \hat\mu <\hat \eta $.  
\end{abstract}

\maketitle

\section{Introduction}
 The rotation law in the infinite (in axial direction) Taylor-Couette flow is
$
\Om(R) = a+b/{R^2},
$
where $a$ and $b$ are  related to the angular
velocities $\Om_{\rm in}$ and $\Om_{\rm out}$ of
the inner and the outer cylinders. The rotation laws  with $a=0$ are  called as located  at the Rayleigh line. With $R_{\rm in}$ and 
$R_{\rm out}$  as the radii
of the  cylinders the parameters 
$\hat\mu={\Om_{\rm out}}/{\Om_{\rm in}}$ and 
$
\hat\eta={R_{\rm in}}/{R_{\rm out}}$ of the flow are defined.
According to the Rayleigh criterion  the ideal flow is hydrodynamically stable  when
the specific angular momentum increases outwards, i.e.
$
\hat\mu > \hat\eta^2
$
or $a>0$. The viscosity, however,  stabilizes the flow so that  it becomes unstable only if 
 the inner cylinder rotates sufficiently fast. 
\begin{figure}[htb]
\mbox{
\includegraphics[scale=0.30]{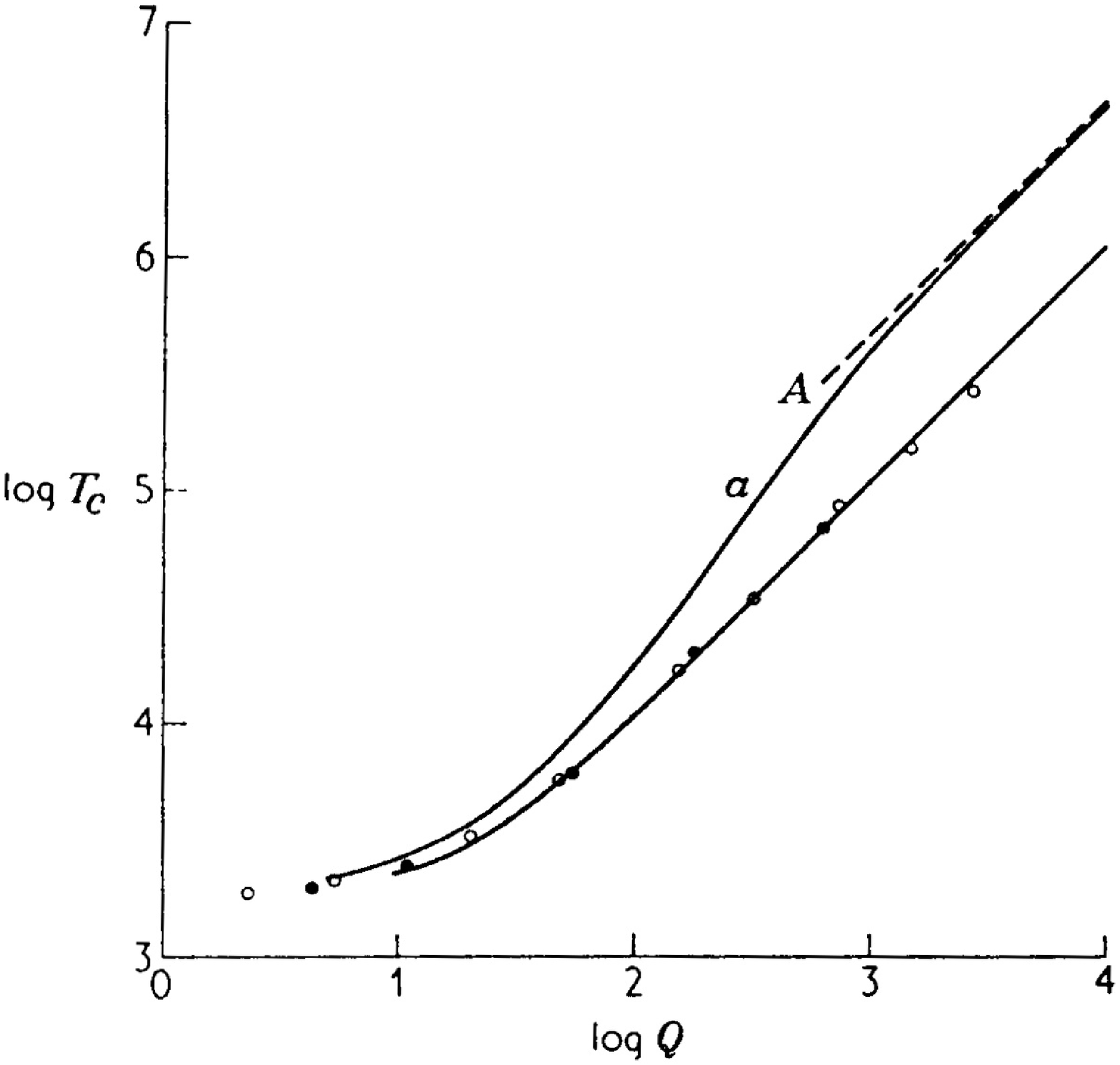}
\hspace{1.2cm}
\includegraphics[scale=0.30]{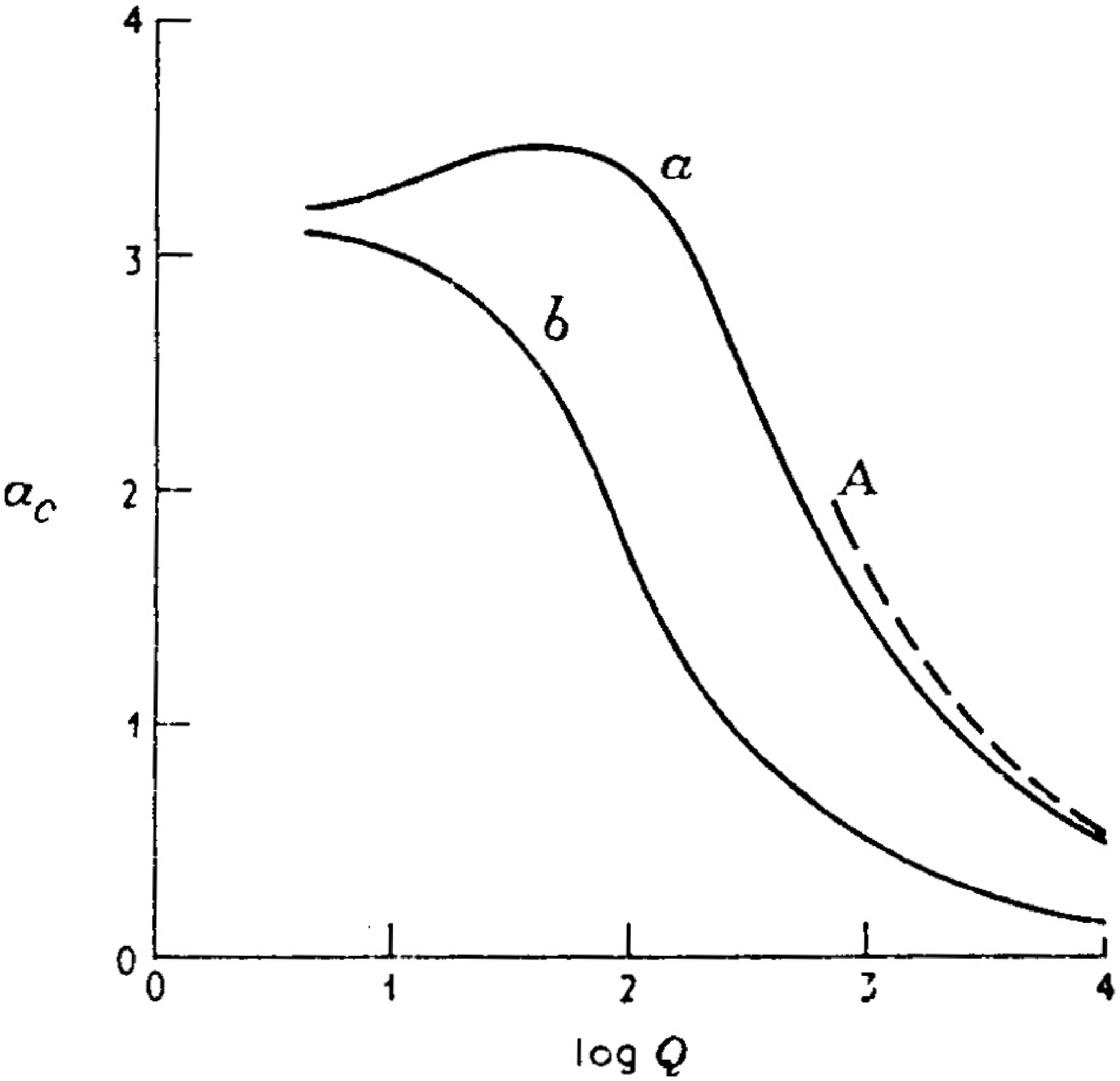}}
\caption{\label{fig71c} Early  results for 
small gap and small Pm (Chandrasekhar 1961). {Left}: The influence of the magnetic field 
(Q here  the Hartmann number) on the critical Taylor number. Magnetic 
fields suppress the instability. {Right}: The same for the wave number. 
The magnetic field elongates the Taylor vortices  in the vertical direction.  Boundary 
conditions: (a) conducting walls, (b) insulating walls.}
\end{figure}

If the fluid is electrically conducting and an axial magnetic field is applied then after the old results the critical
Reynolds number (for the inner rotating cylinder) grows with growing  magnetic field. Figure~\ref{fig71c}
shows the theoretical results of
Chandrasekhar (1961) for narrow gaps and very small magnetic Prandtl numbers, ${\rm Pm}=\nu/\eta$,  together with the experimental data for Mercury  of Donnelly \& Ozima (1960).
Theory and observations are in nearly perfect agreement with no indication of any
magnetic-induced instability.  The magnetic Prandtl number under laboratory
conditions is really very small ($10^{-5}$ for liquid sodium).

\begin{figure}[htb]
\includegraphics[height=7cm,width=8cm]{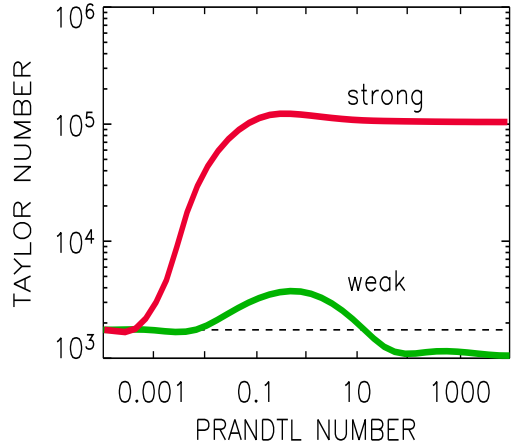}
\caption{\label{fig71b} Critical Taylor number as a 
function of the magnetic Prandtl number for strong  and weak 
 magnetic fields within the small-gap approximation. For weak
 fields and large magnetic Prandtl numbers the system is
subcritical. For $B_0=0$ (or, what is the same, for ${\rm Pm}=0$) is  Ta$_{\rm crit}= 1750$.  (Kurzweg (1963).}
\end{figure}

Within the
small-gap approximation but with free Pm  Kurzweg (1963) found
that for {\em weak}
magnetic fields and sufficiently large magnetic Prandtl number the critical
Taylor number becomes {\em smaller}
than in the hydrodynamic case (Fig.~\ref{fig71b}). If the field is not too strong it can basically play a destabilizing role.  

 Velikhov (1959) originally discovered this magnetic shear-flow instability which  is now called `magnetorotational instability' (MRI).
He found that for 
the ideal hydromagnetic Taylor-Couette flow the Rayleigh criterion for stability changes to 
$
\hat\mu > 1,
$
i.e. only flows with superrotation are still stable.  He  found a growth rate along the 
Rayleigh line (i.e. $a=0$) of $2\Om_{\rm in} \hat\eta$ and a 
critical wave number of 
\beg
k\leq 2\hat \eta \frac{\Om_{\rm in}}{V_{\rm A}},
\label{veli}
\ende 
with $V_{\rm A}$ the Alfv\'en velocity of the given axial field. Only if  $V_{\rm A}$ is {\em smaller} than the shear $-R^2 {\rm d}\Om/{\rm d}R$ the instability works which, therefore, is a weak-field instability. 

The hydrodynamic
Taylor-Couette flow is  stable if its angular momentum increases with radius,
but the hydromagnetic Taylor-Couette flow is only stable if the angular velocity
itself increases with radius. This remains true also for nonideal fluids. The
MRI reduces the critical
Reynolds number for weak magnetic field strengths for hydrodynamically unstable
flow and it destabilizes the otherwise hydrodynamically stable flow for
$
\hat\eta^2 < \hat\mu < 1.
$
The MRI 
exists, however, in hydrodynamically unstable situations
($\hat\mu < \hat\eta^2$) only if Pm is not very
small (Fig.~\ref{fig71b}).

As we shall demonstrate  the magnetic
Reynolds number 
\beg
\rm Rm = Re\ Pm
\label{rm}
\ende
 controls the instability.
Because of the high value of $\eta$ for liquid metals (exceeding  1000~cm$^2$/s), 
 it is  not easy to reach magnetic Reynolds numbers of the required order of $1\dots 10$. 
This is the basic  reason why the MRI has not yet been  observed experimentally 
in the laboratory.

\section{Magnetorotational instability (MRI)}\label{s83}

\begin{figure}
\mbox
{
\includegraphics[height=7cm,width=7.5cm]{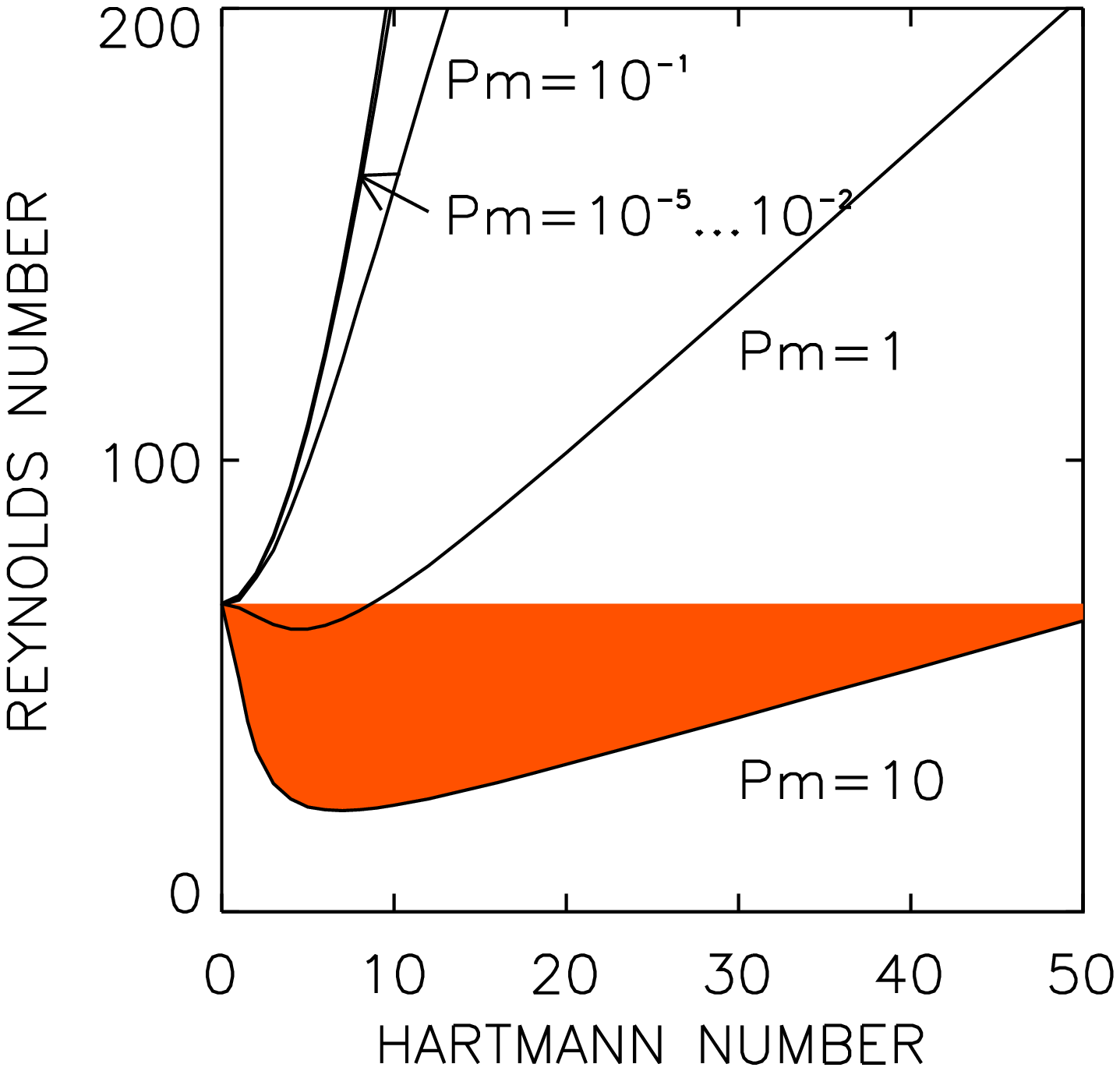}
\includegraphics[height=7cm,width=7.5cm]{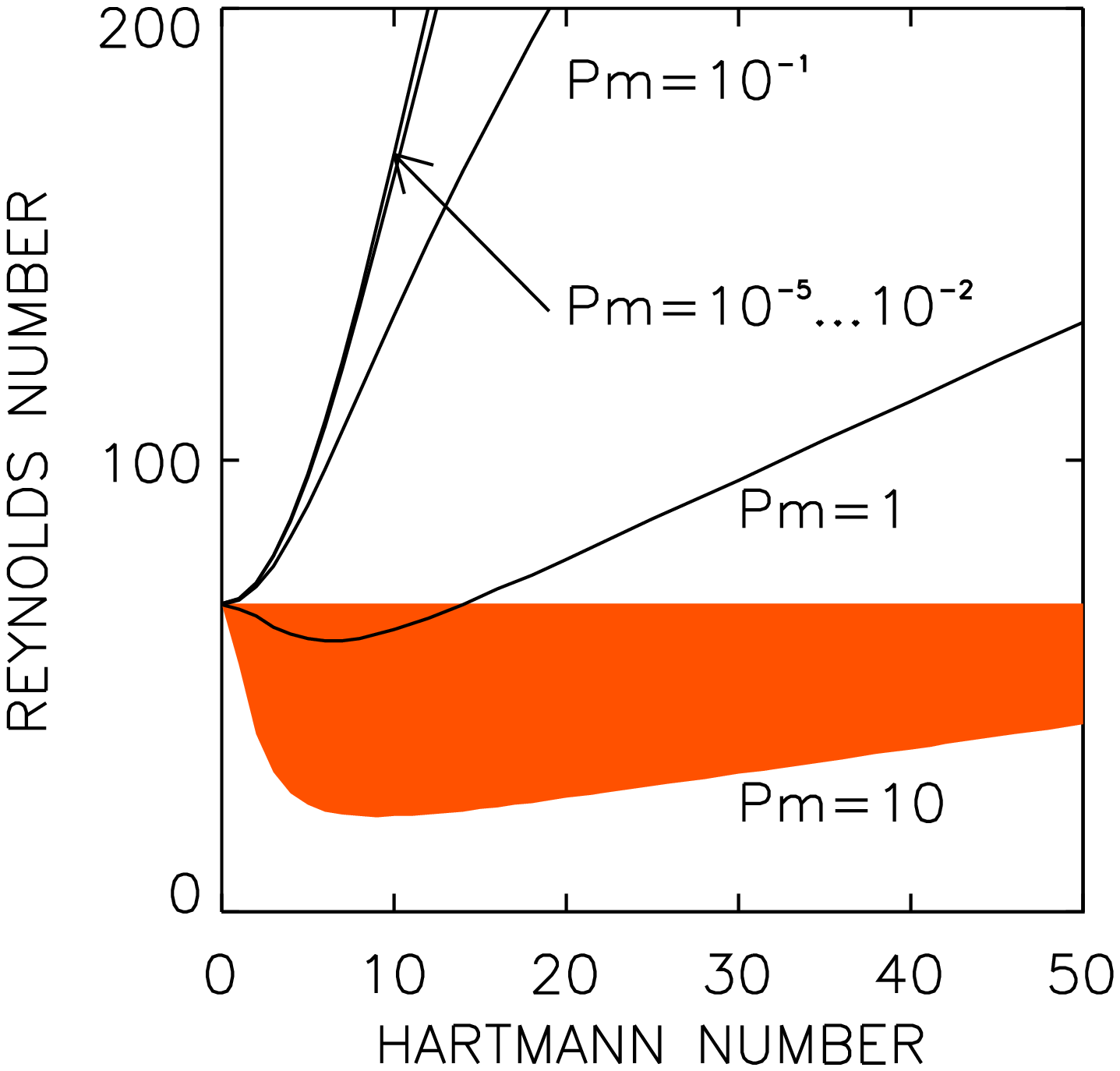}}
\caption{\label{fig73a} Bifurcation diagram  for axisymmetric modes 
with resting outer cylinder of conducting material ({left}) and vacuum
({right}). Shaded areas denote
subcritical excitation  by the axial 
magnetic field. $\hat\eta=0.5$. From  R\"udiger, Schultz \& Shalybkov (2003).}
\end{figure}

The Reynolds number is usually defined as ${\rm Re} = R_{\rm in}(R_{\rm out}-R_{\rm in})\Omega_{\rm in}/\nu$. The amplitude of the external magnetic field $B_0$ is expressed with the Hartmann number ${\rm Ha} = B_0 \sqrt{R_{\rm in}(R_{\rm out}-R_{\rm in})/(\mu_0\rho\nu\eta)}$. 
The  values of the Reynolds numbers above which the flow
becomes unstable depend on the vertical wave number. They
have a minimum at some wave number for fixed other parameters.
This minimum value is called the critical Reynolds number.
 Figure~\ref{fig73a} shows the neutral stability for axisymmetric modes for
containers with both conducting and insulating walls with resting outer cylinder
and for 
fluids of various magnetic
Prandtl number. 
 ${\rm Re}= 68$ is the classical hydrodynamic solution for resting outer cylinder and  $\hat\eta=0.5$. There is a strong difference of
the bifurcation lines for ${\rm Pm} \gsim 1$ (high conductivity) and ${\rm Pm}
<1$ (low conductivity). For
fluids with low electrical conductivity the magnetic field only
suppresses the instability so that all the critical Reynolds numbers strongly exceed
the  value 68.

For small magnetic Prandtl number the stability lines hardly
differ.
The opposite is true for ${\rm Pm} \gsim 1$. In Fig.~\ref{fig73a} 
the resulting critical Reynolds
numbers Re are smaller than 68. The magnetic fields with small Hartmann
numbers support the instability  rather than to suppress it. This
effect becomes more effective for increasing Pm but it vanishes for 
stronger magnetic fields. The 
MRI only exists for weak magnetic fields and high enough electrical conductivity
and/or microscopic   viscosity.

\begin{figure}
\mbox
{
\includegraphics[height=8cm,width=7.8cm]{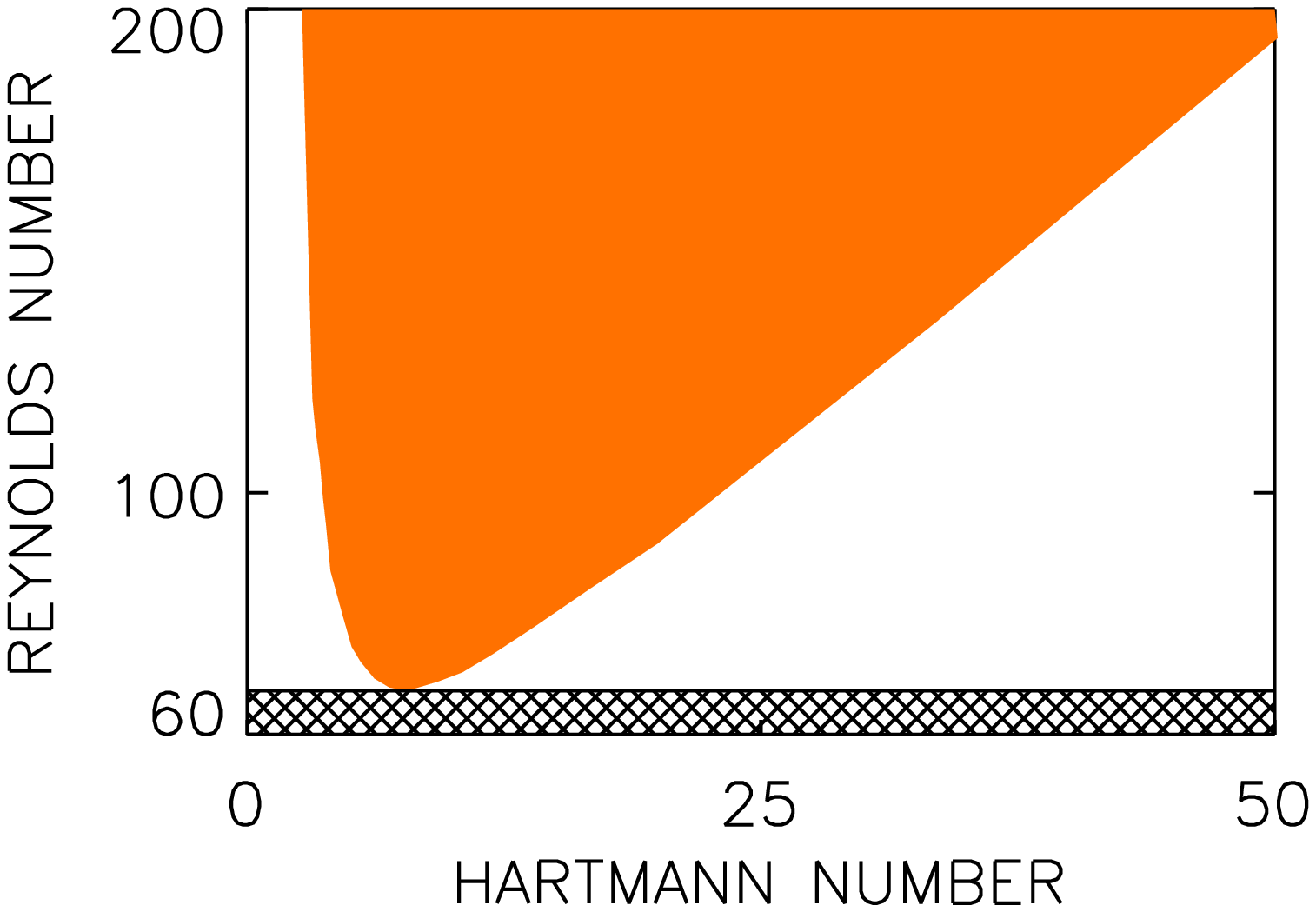}\hfill
\includegraphics[height=8cm,width=7.8cm]{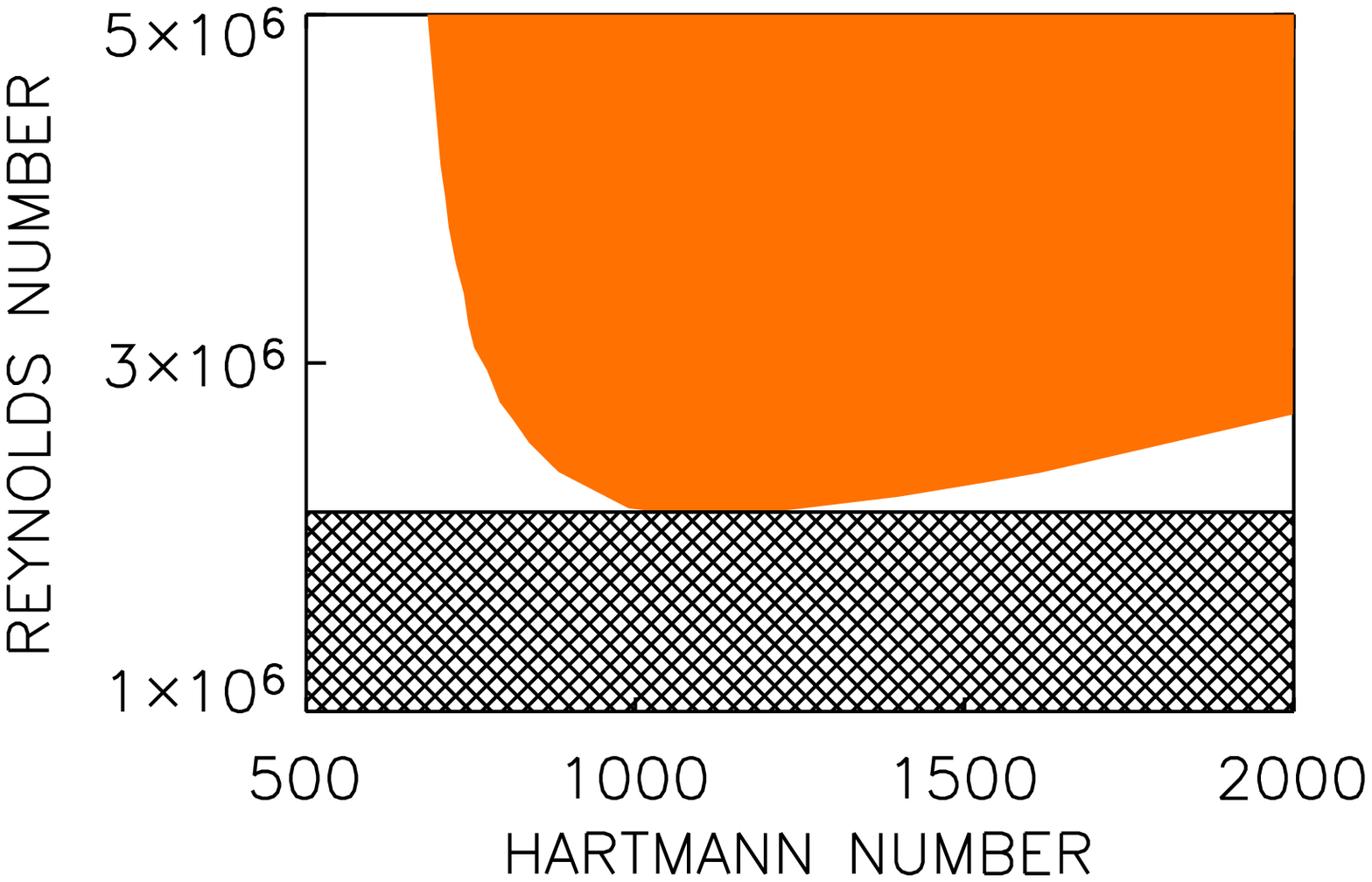}}
\caption{\label{fig73aa} Marginal stability lines for axisymmetric modes in containers 
with rotating outer cylinder  of conducting material for $\rm
Pm=1$ ({ left}) and $\rm Pm= 10^{-5}$ ({right}). $\hat \eta=0.5, \hat\mu
=0.33$. The instability domain is colored by grey and fluids in the
cross-hatched area are always stable.}
\end{figure}

\begin{figure}
\includegraphics[scale=0.7]{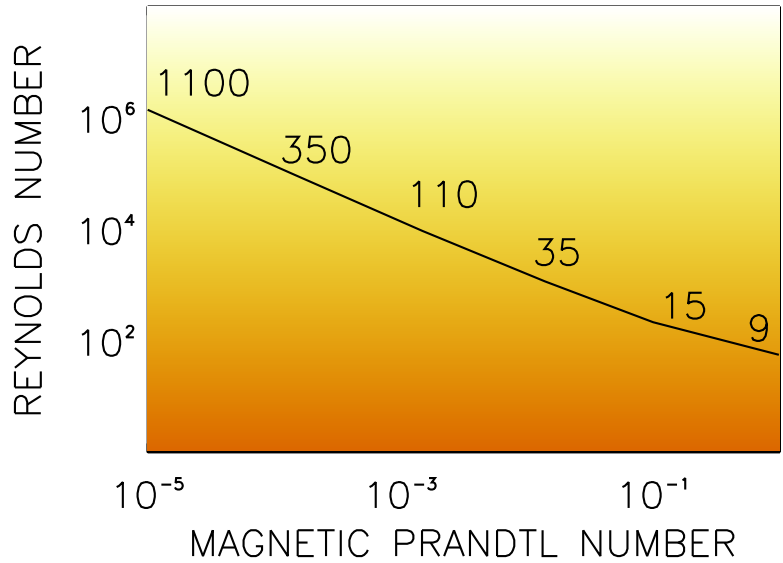}
\caption{The critical Reynolds numbers vs. magnetic
 Prandtl numbers 
marked with those Hartmann numbers where the Reynolds number is 
minimum. $\hat \eta=0.5$, $\hat \mu = 0.33$.}
\label{fig73bb}
\end{figure}

Now  the outer cylinder may rotate so fast that the
rotation law no longer fulfills the Rayleigh criterion and a solution 
for ${\rm Ha}=0$  does not exist. The nonmagnetic eigenvalue along the vertical 
axis moves to infinity but {\em a minimum remains}. 
Figure~\ref{fig73aa}  presents the results for
$\rm Pm=1$ and $\rm Pm= 10^{-5}$. There are always minima of Re for 
certain Hartmann numbers (R\"udiger \& Shalybkov 2002). The minima and the critical Hartmann numbers 
increase for decreasing magnetic Prandtl numbers.
For $\hat\eta=0.5$ and $\hat\mu=0.33$ the  critical Reynolds numbers together 
with the critical Hartmann numbers are plotted in Fig.~\ref{fig73bb}. 
Table~\ref{tab73a} gives the exact coordinates of the absolute minima for 
experiments with rotating outer cylinder and  for $\rm Pm= 10^{-5}$. They are characterized by magnetic Reynolds numbers of order  10, very similar to 
the values of the existing dynamo experiments.
 
\begin{table}
\caption{Coordinates of the absolute minima in 
Fig.~\ref{fig73aa} for rotating outer cylinder with $\hat \mu=0.33$, 
$\hat \eta = 0.5$ and  $\rm Pm= 10^{-5}$ .}
\begin{tabular}{@{}lll@{}}
\hline
&&\\[-1.5ex]
 & insulating walls & conducting walls\\[1ex]
\hline
&& \\[-1.5ex]
Reynolds  number & $1.42 \cdot 10^6$ & $2.13 \cdot 10^6$\\
mag. Reynolds  number & 14 & 21\\
Hartmann number & $1400$ & $1100$\\
Lundquist number & 4.42 & 3.47\\[1ex]
\hline
\end{tabular}
\label{tab73a}
\end{table}

A container with  an outer
radius of 22~cm  (and an inner radius of 11~cm)  filled with liquid
sodium
requires a {\em rotation of about 19 Hz} in order to find the MRI. The required magnetic field is about 1400 Gauss.

The results for containers with {\em conducting} walls are also given in 
Table~\ref{tab73a}.  
The minimal Reynolds numbers  are
higher than for insulating cylinder walls. The influence of the 
boundary conditions is thus not  small\footnote{for not too wide gaps}.

\begin{figure}[ht]
\mbox{
\includegraphics[height=6.5cm,width=6.9cm]{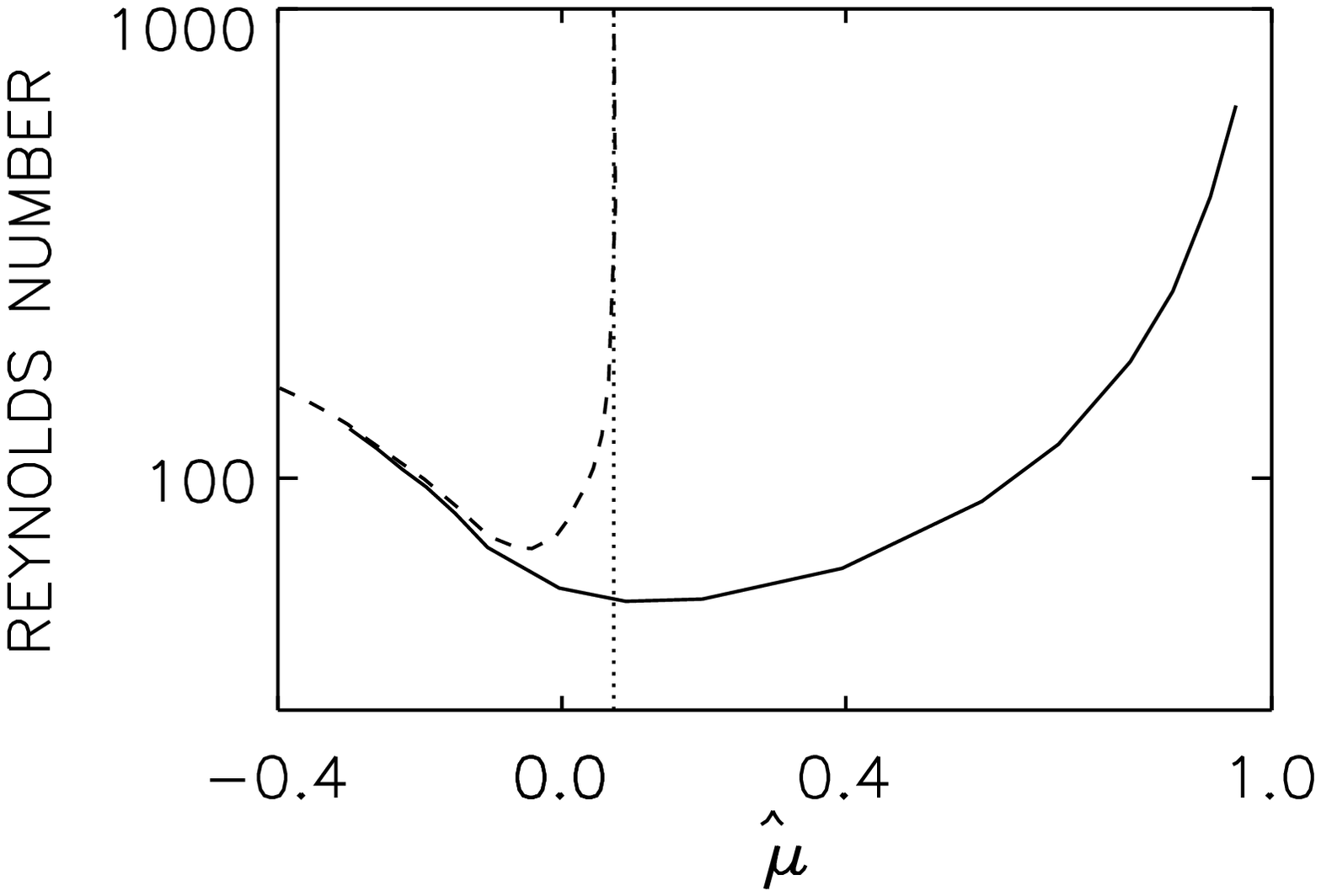}\hfill
\includegraphics[height=6.5cm,width=6.9cm]{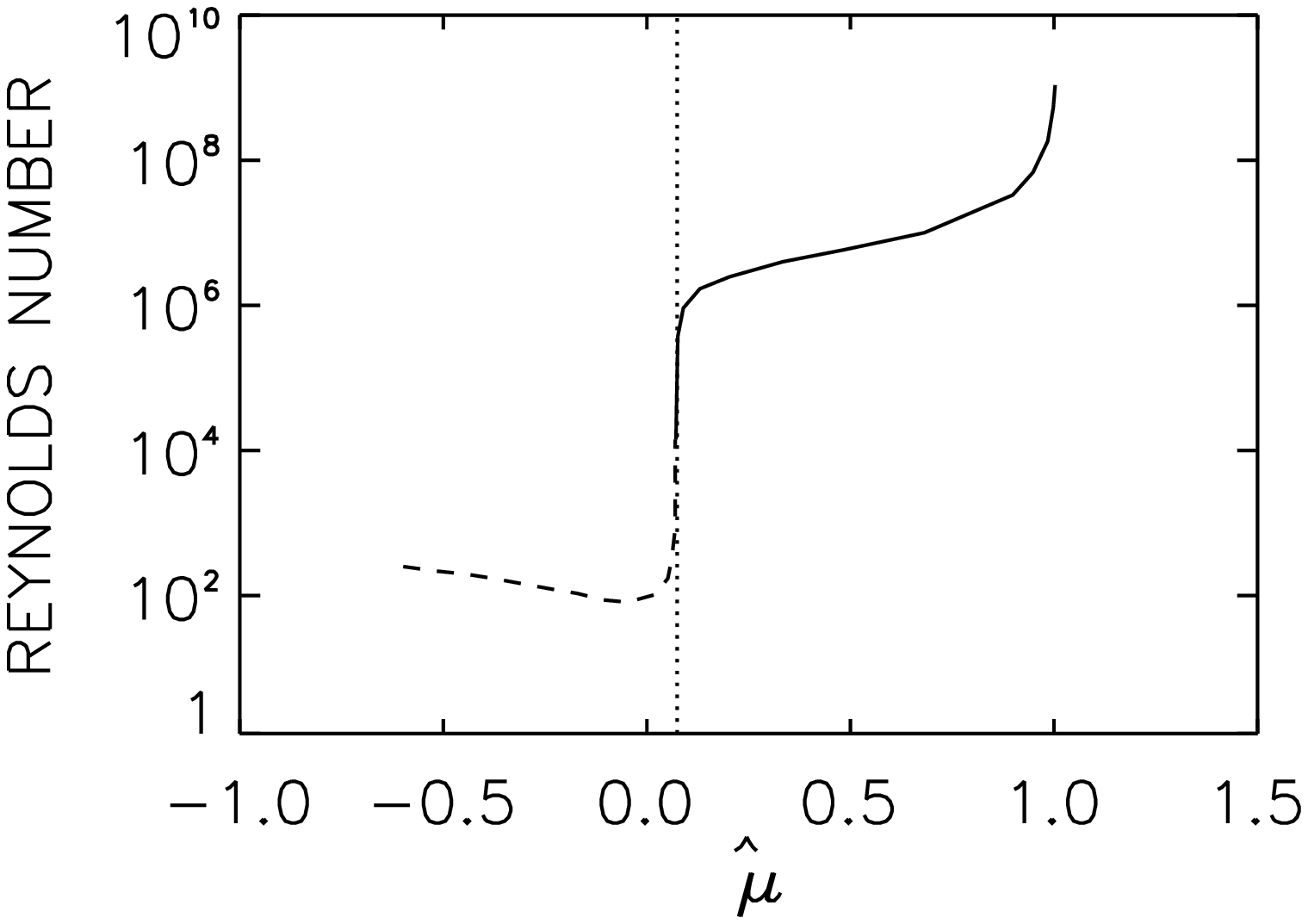}}
\caption{Critical Reynolds numbers for the Taylor-Couette flow vs. $\hat\mu$
for $\hat\eta=0.27$ and  ${\rm Pm}=1$ ({left}) and  ${\rm Pm}=10^{-5}$ ({right}). From all $\hat\eta$, $\hat\eta=0.27$ yields the lowest minimum. 
The  curve for the hydrodynamic  instability (${\rm Ha}=0$) is {dashed} and  the 
hydromagnetic   curve (${\rm Ha}>0$)  is {solid}.  The  {dotted lines} denote the location 
of  $a=0$.}
\label{fig73b}
\end{figure}

There is a particular scaling for the special case of $a=0$, i.e. for $\hat\mu =\hat\eta^2$. One finds that the quantities $u_R, u_z, B_R$ and $B_z$
scale
as ${\rm Pm}^{-1/2}$ while $u_\phi, B_\phi, k$ and  Ha scale as ${\rm Pm}^0$.
The Reynolds number for the axisymmetric modes  scales as
$
{\rm Re} \propto {\rm Pm}^{-1/2}$ (Willis \& Barenghi 2002).
The scaling does not depend on the boundary
conditions. However,
  for $a > 0$  
 the much steeper
scaling
$
{\rm Re} \propto {\rm Pm}^{-1}
$ results , 
leading to the surprisingly simple relation
\beg
{\rm Rm}\simeq {\rm const.}
\label{73.3}
\ende
for the magnetic Reynolds number Rm (Fig.~\ref{fig73bb}). For the Lundquist  number ${\rm S} = {\rm Ha} \sqrt{{\rm Pm}}$ of the
characteristic minima we find
\beg
{\rm S} \simeq {\rm const.}
\label{73.4}
\ende
For small magnetic Prandtl numbers the exact value of 
 the microscopic viscosity is thus {\em not} relevant for the 
 instability. In consequence, the corresponding Reynolds numbers for the MRI seem to differ by 2 orders of magnitude, i.e.  
10$^4$  and 10$^6$ at and shortly beyond the Rayleigh line.
 Figure~\ref{fig73b} 
shows
the behavior close to $\hat\mu = \hat\eta^2$.
There is a vertical jump  within an extremely small
interval of $\hat \mu$. This sharp transition does not exist for ${\rm Pm}=1$;
it only exists for very small
values of Pm.  Even the smallest
deviation from the condition $\hat\mu = \hat\eta^2$ would drastically change the
excitation condition.

\begin{figure}[htb]
\includegraphics[width=14cm,height=7cm]{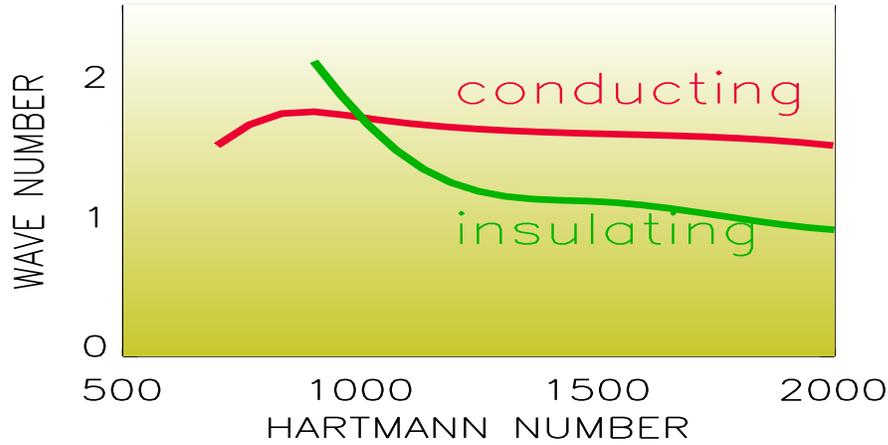}
\caption{\label{fig73f}  Wave 
numbers  for which the Reynolds number is minimum  for conducting walls and insulating 
walls.  $\hat\eta=0.5$, $\hat \mu=0.33$,    
${\rm Pm}=10^{-5}$. Note how exact the Velikhov condition (\ref{veli}) is fulfilled.  From R\"udiger \& Shalybkov (2002).}
\end{figure}

For $\hat\eta\simeq 0.5$ the  Velikhov condition (\ref{veli}) can also be written as $k R_{\rm in} \simeq {\rm Rm/S}$. The vertical extension $\delta z$ of a Taylor vortex is 
given by 
\beg
\frac{\delta z}{R_{\rm out} - R_{\rm in}} \simeq \frac{\pi}{k}
\sqrt{\frac{\hat \eta}{1-\hat\eta}}.
\label{73.7}
\ende
The dimensionless vertical wave numbers $k$  associated with
 the critical Reynolds numbers  are given 
in Fig.~\ref{fig73f}. 
 For hydrodynamically unstable flows we have
$\delta z \simeq R_{\rm out} - R_{\rm in}$ for small magnetic fields
(${\rm Ha}\simeq0$).
The cell therefore has the same vertical extent as it has in 
radius.

The  magnetic fields deforms the Taylor vortices. 
The deformation consists in a elongation of the cell in the 
vertical direction. The wave number is thus expected to 
become smaller and smaller for increasing
magnetic field. This is indeed true for  ${\rm Pm}\simeq1$, but for smaller Pm
the situation is more complicated.

The cell size is minimum for the critical Reynolds number for all calculated
examples of hydrodynamically stable flows with a conducting boundary.
This  is not true, however,
for containers with insulating walls,  for which the  cell size  grows with
increasing magnetic field. 
For experiments with the critical Reynolds numbers the vertical cell
size is generally 2--3 times larger than the radial one.
The smaller the
magnetic Prandtl number the longer are the cells in the vertical direction.
The influence of boundary conditions on the cell size disappears 
 for sufficiently wide gaps.

\begin{figure}[htb]
\includegraphics[height=6cm,width=10cm]{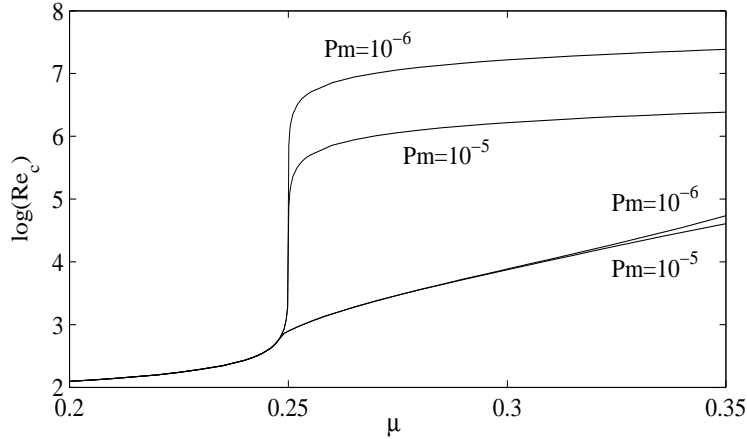}
\caption{Critical Reynolds numbers for marginal stability of  axisymmetric modes in containers ($\hat \eta=0.5$)
with purely axial magnetic field (upper curves) and with extra (current-free) toroidal field (lower curves) for  $\rm Pm= 10^{-5}$ and $\rm Pm= 10^{-6}$ as a function of $\hat\mu$. Note the extremely weak Prandtl number dependence of the lower curves and the strong reduction of the critical Reynolds numbers. From Hollerbach \& R\"udiger (2005). }
\label{toro}
\end{figure}

The given solutions of marginal stability of axisymmetric modes are stationary. One could ask for the character of the solutions if an extra  toroidal magnetic field  is applied (Hollerbach \& R\"udiger 2005). We know that current-free toroidal fields alone do not change the stability of Taylor-Couette flow (Velikhov 1959). If axial fields {\em and} current-free toroidal fields of the same order  exist in the container, however, completely  new solutions appear and they are oscillating  
(Fig.~\ref{toro}).  The real parts of the eigenfrequencies  are positive so that stationary modes do not longer exist. The Reynolds number of the flow and the  Hartmann number of the axial field which are necessary to induce the MRI instability are
strongly reduced by the toroidal field. In Table~\ref{tab1} 
  characteristic numbers from numerical experiments with small magnetic Prandtl number 
and for $\hat\mu=0.3$ are given. 

\begin{table}
\caption{\label{tab1} Characteristic values for $\hat \eta=0.5$, $\hat\mu=0.3$, $\rm Pm=10^{-5}$. $\beta$ describes the amplitude ratio of the toroidal and the axial magnetic field.}
\medskip
\begin{tabular}{l|cccc}
\hline
  $\beta$  & Re & Ha & $k$ & $\Re(\omega)$\\[0.5ex]
\hline
0 & $1.6 \cdot 10^6$ &  850 & 1.7 & 0\\[0.5ex]
1 &$5.8 \cdot 10^5$ &383& 0.47 & 0.04\\[0.5ex]
 \end{tabular}
 \end{table}

\section{Shear-Hall instability (SHI)}
In the presence of the Hall effect there is even the possibility that differential rotation with $\Omega$ increasing outwards (i.e. $\partial \Omega/\partial R>0$) becomes unstable. The Hall effect 
destabilizes flows with $\hat\mu >1$ 
for which so far no other instability is known. The dispersion relation derived from the equation system with an induction equation including Hall effect leads to the instability condition ${\rm Rb}\ \partial \Omega/\partial R<0$ with Rb as the ratio of  the Ohmic time scale and the Hall time scale. It linearly depends on the electrical conductivity and the magnetic field (see R\"udiger \& Hollerbach 2004). For magnetic fields parallel to the rotation axis it might be positive and v.v. The result  depends on the sign of the Hall resistivity; here 
the positive Hall resistivity  is used.
\begin{figure}[htb]
\mbox{
\includegraphics[width=7.3cm,height=6cm]{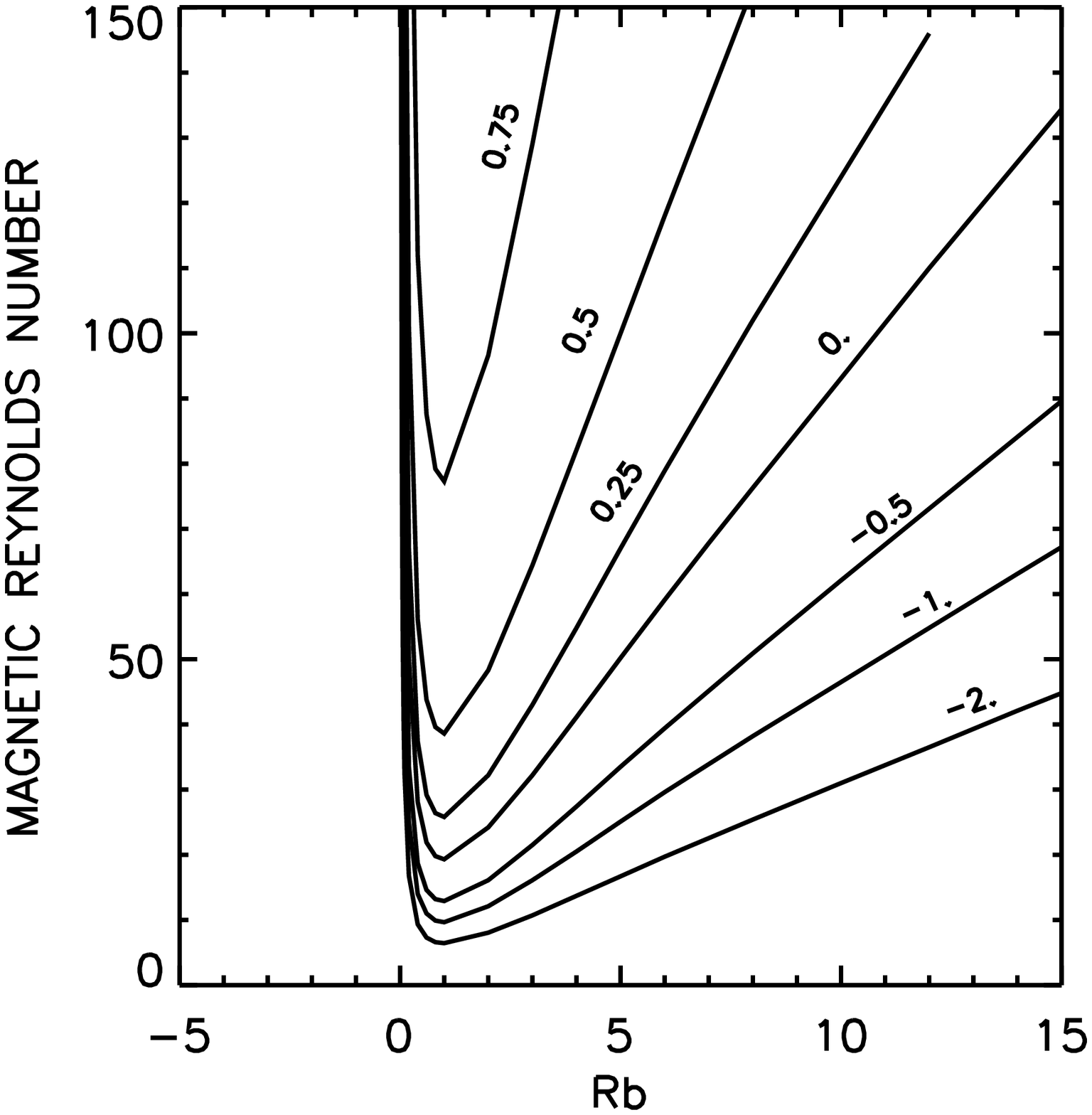}\hfill
\includegraphics[width=7.3cm,height=6cm]{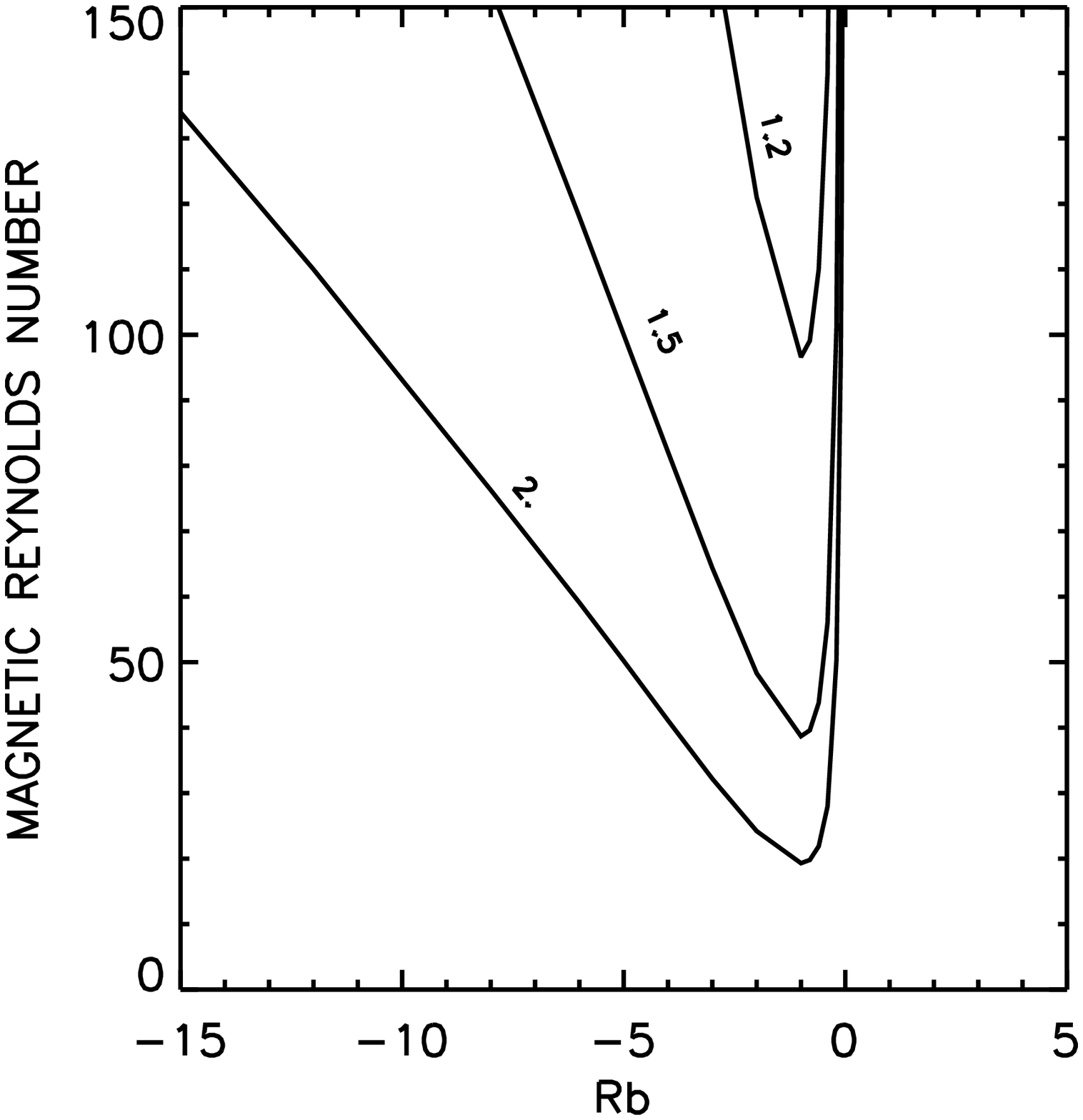}}
\caption{The shear-Hall instability. Only the induction equation is solved for Taylor-Couette flows under the presence of the Hall effect. Left: magnetic field positive, shear negative. Right: magnetic field negative, shear positive. The lines of marginal instability are labeled with the actual value of $\hat\mu$. }\label{hall1}
\end{figure}

Figure~\ref{hall1} illustrates the instability
for   a container with again
$\hat\eta=0.5$. The flow with $\hat\mu<1$  is unstable  for positive Rb. Flows with  $\hat\mu>1$ are unstable for negative Rb, i.e. if angular velocity and magnetic field have opposite 
directions\footnote{For
negative Hall resistivity the orientation is opposite}.
The fact that the Hall effect destabilizes flows with the angular velocity 
increasing outwards was first described by Balbus and Terquem (2001). 

Figure~\ref{hall1} may only serve as an illustration. In particular for $\hat\mu<1$ the situation is  more complicated as simultaneously the MRI is active. One finds more details in the original paper of 
R\"udiger \& Shalybkov (2004) about the overall impact of the Hall effect on the Taylor-Couette flow. That the {\em sign} of the magnetic field has such an important influence on stability or instability of flows (e.g. Kepler flows) must have important consequences for astrophysical applications.

\begin{figure}[ht]
\mbox{
\includegraphics[width=8cm,height=6.0cm]{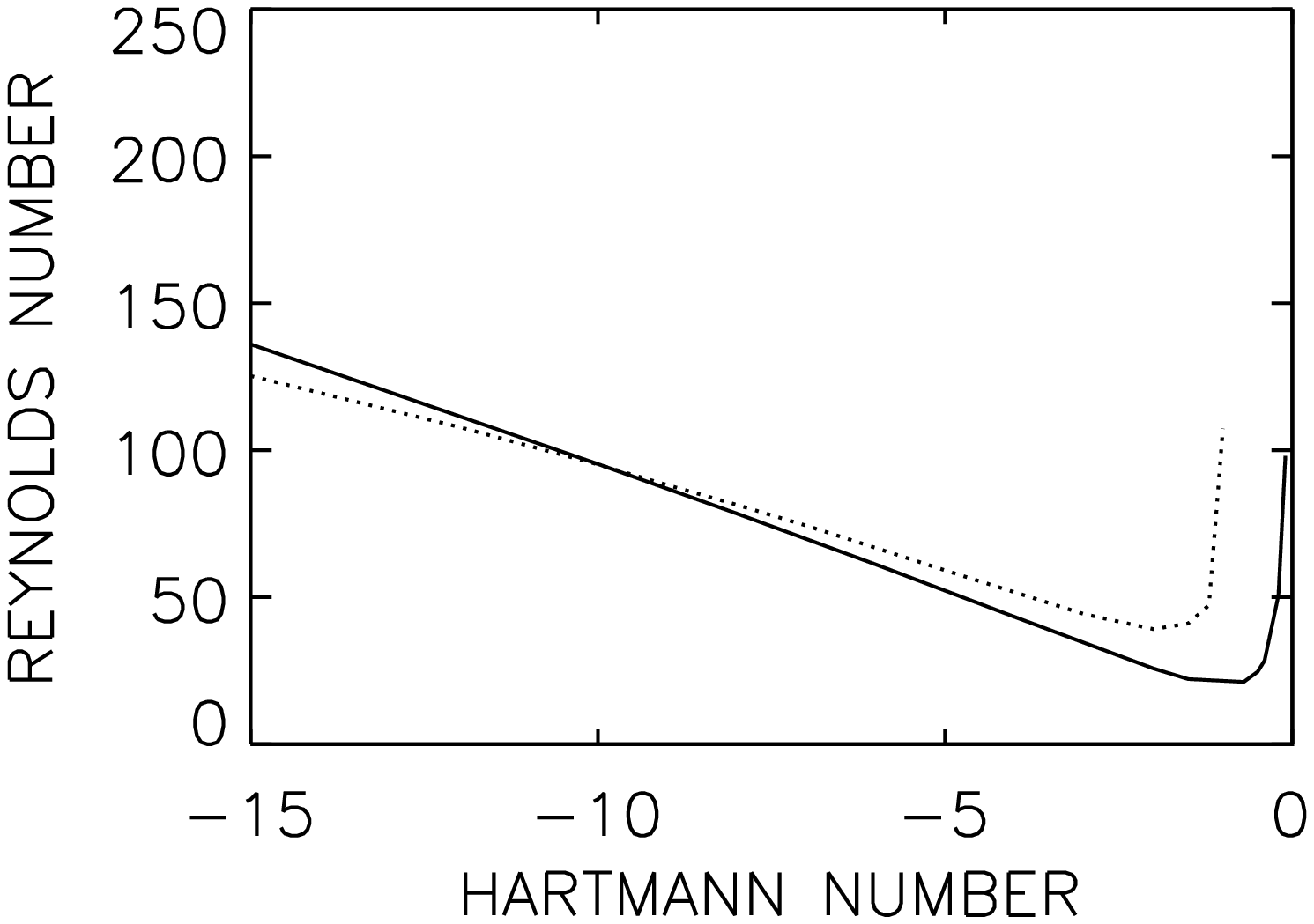}\hfill
\includegraphics[width=8cm,height=6.0cm]{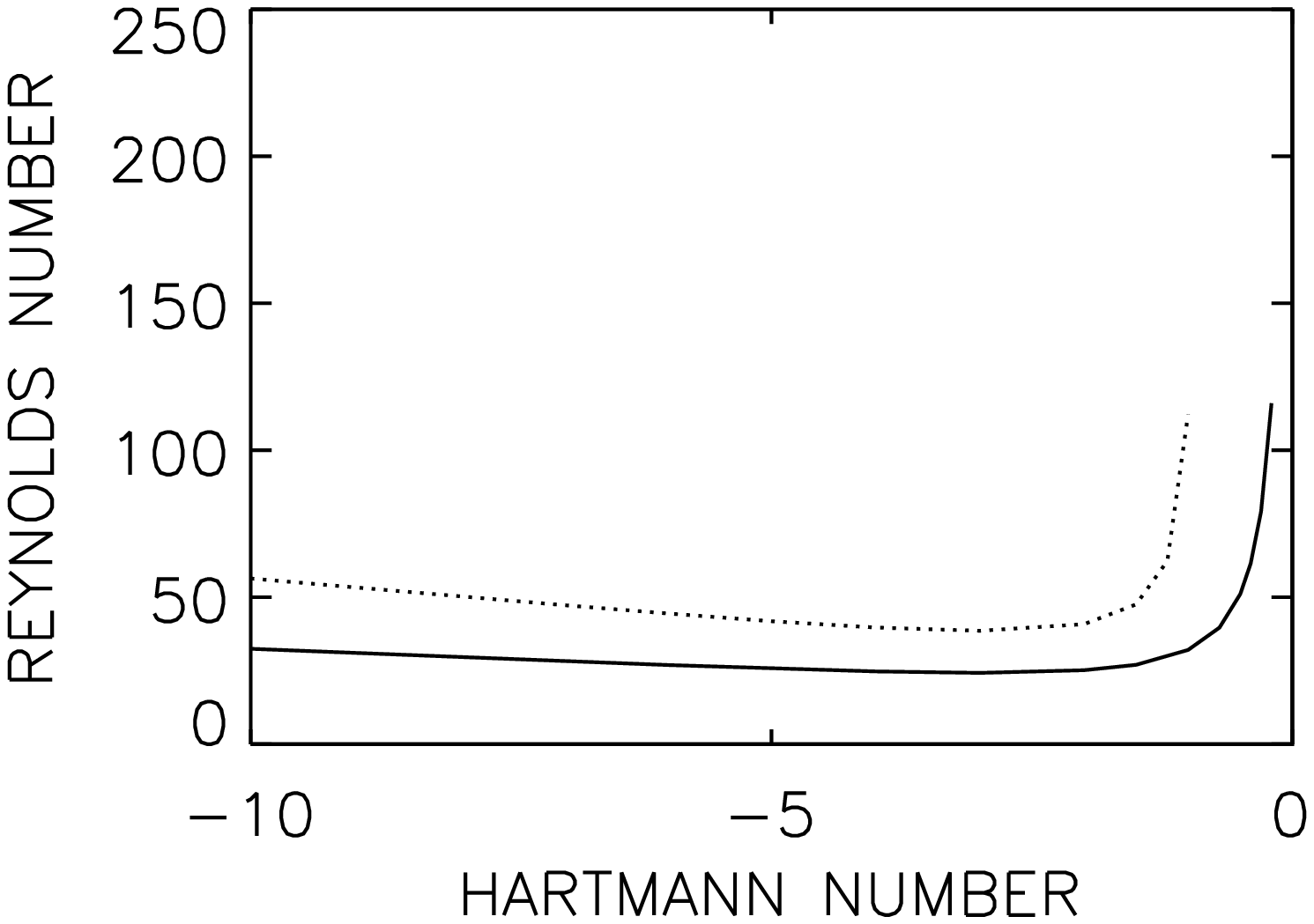}} 
\caption{ The  lines of marginal instability   for $\hat\mu=2$ with axisymmetric ($m=0$, solid line) and 
nonaxisymmetric ($m=1$, dotted line) modes. Left: conducting walls, note the crossover of both lines. Right: insulating walls, no crossovers. Here $\rm Ha=Rb$.
}
\label{nonaxi}
\end{figure}

It is also important to know the  nonaxisymmetric modes. After
the Cow\-ling theorem only nonaxisymmetric modes can be maintained by a dynamo
process. We  have discussed the appearance of
nonaxisymmetric modes for the magnetic Taylor-Couette flow with negative
shear (Shalybkov, R\"udiger \& Schultz 2002). The common result was
that the lines of marginal stability for $m=0$ and $m=1$ have a very
different behavior for different electrical boundary conditions.
One finds crossovers of the stability lines for containers with conducting
cylinder walls, and one never finds crossovers for containers with vacuum
boundary conditions. The same happens  for the Hall instability for
magnetic Taylor-Couette flows with $\hat\mu > 1$. In
Fig.~\ref{nonaxi} the lines for both axisymmetric and nonaxisymmetric modes
are given for conducting boundary conditions and for vacuum boundary conditions. Indeed, a crossover of the lines only
exists for conducting cylinder walls. As usual, in the minimum the $m=0$ mode
dominates but for stronger magnetic fields the mode with $m=1$ dominates.

\section{STRATOROTATIONAL INSTABILITY (SRI)}
The hydrodynamic Taylor-Couette flow stability is now discussed  under the presence of
vertical density stratifications. 
Taylor-Couette flows with a stable axial density stratification were 
often studied with the conclusion  that a stable 
stratification stabilizes the flow.
Withjack \& Chen (1974), however,  reported new experiments with  the Rayleigh line  at $\hat\mu=0.04$. For a slight negative density gradient   disturbances of spiral form were observed for $\hat\mu=0.073$, i.e. clearly beyond the Rayleigh line. 
Recently, Yavneh, McWilliams \& Molemaker (2001) find that
the sufficient condition for stability against nonaxisymmetric disturbances is
the same as the condition for MRI.
Their numerical results  demonstrated that certain instabilities exist also 
for finite viscosity. 

\begin{figure}[htb]
\includegraphics[width=6cm,height=6.5cm]{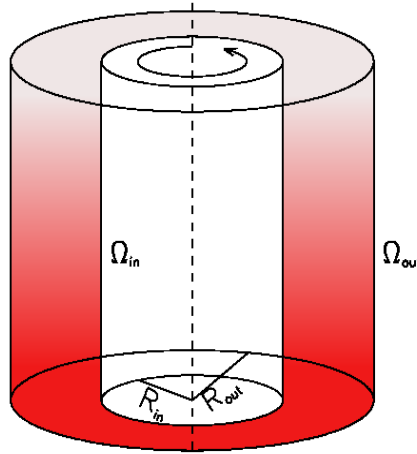}
\caption{Taylor-Couette flow geometry with axial density stratification }
\label{shal1}
\end{figure}

One has to look  for the basic state with prescribed  
velocity profile and given density
vertical stratification $\rho=\rho(z)$.
The initial system  then can only exist if 
\beg
R\Omega^2 \frac{\partial \rho}{\partial z}
+g\frac{\partial \rho}{\partial R} =0.
\label{cond1}
\ende
is fulfilled. Under the centrifugal force 
the purely vertical stratification at the beginning transforms to a
mixed vertical and radial stratification.
This behavior  strongly complicates the problem. 

In real experiments the initial 
vertical stratification is so small that 
$ R^2\Omega \ll g$.
Then after (\ref{cond1}) the radial stratification is also small.
Let us, therefore, consider the `small-stratification case' with
$
\rho=\rho_0+\rho_1(R,z)$ with $\rho_1 \ll \rho_0$,
where $\rho_0$ is the uniform background density and condition
(\ref{cond1}) is fulfilled at zero order.
If only  the terms of
the zero  order are considered then the system takes the Boussinesq form.
$N$ is the vertical Brunt-V\"ais\"al\"a frequency $
N^2=-(g/\rho_0)\partial \rho_1/\partial z
$ and ${\rm Fr}= \Omega_{\rm in}/N$ is the Froude number.

Here we suppose 
$\partial \rho_1/\partial z={\rm{const}}$ so that 
 the coefficients of the system  only depend on the
radial coordinate  and  a normal mode expansion
$
~{\rm{exp}}({\rm{i}}(m\phi+kz+\omega t))
$
is possible. As usual, only those wave numbers are considered which belong to the smallest Reynolds numbers.
We have also checked the existence of the transition from stable to unstable
state for some arbitrary points.
\begin{figure}[htb]
\mbox{
\includegraphics[width=7.4cm,height=7.0cm]{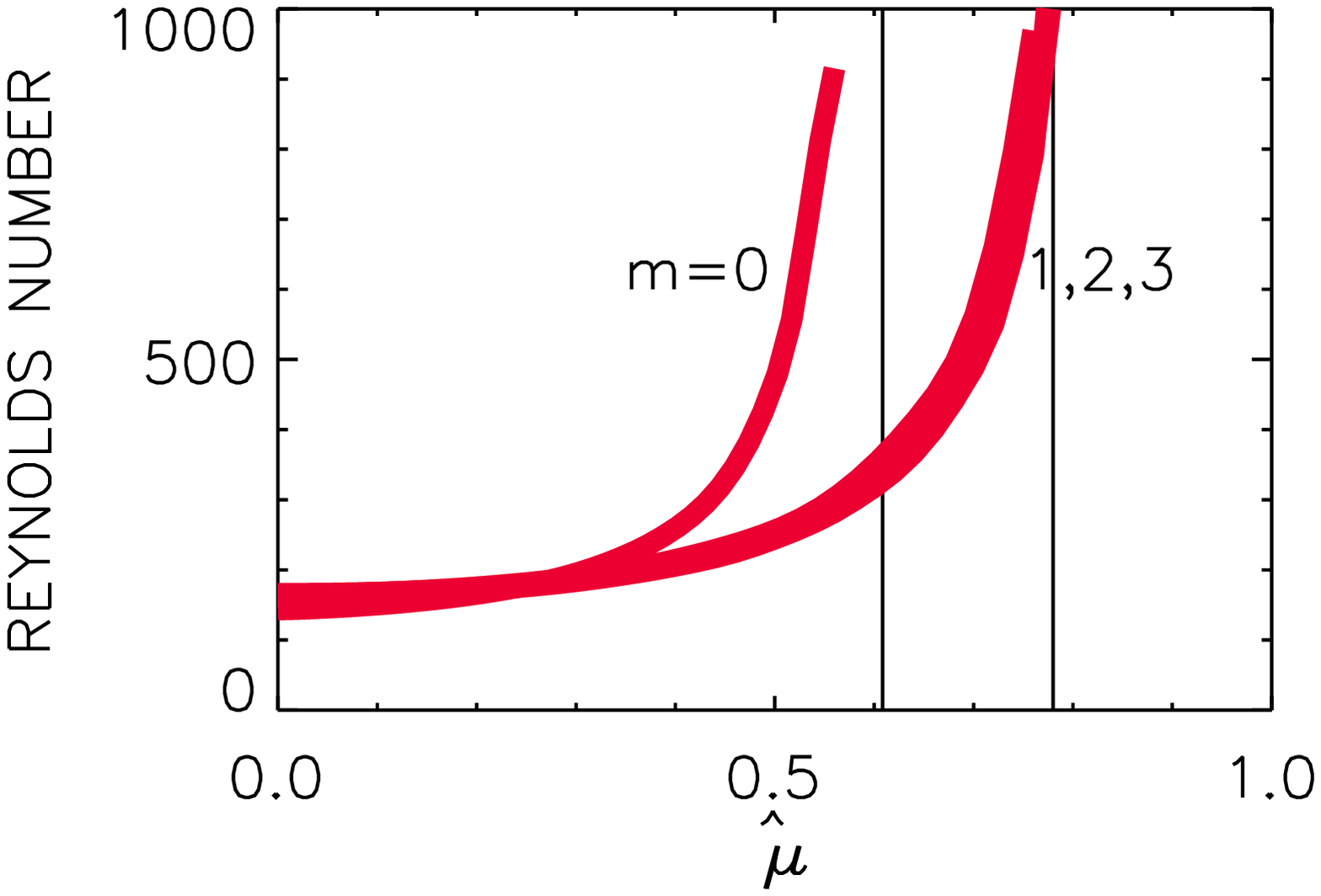} \hfill
\includegraphics[width=7.4cm,height=7.0cm]{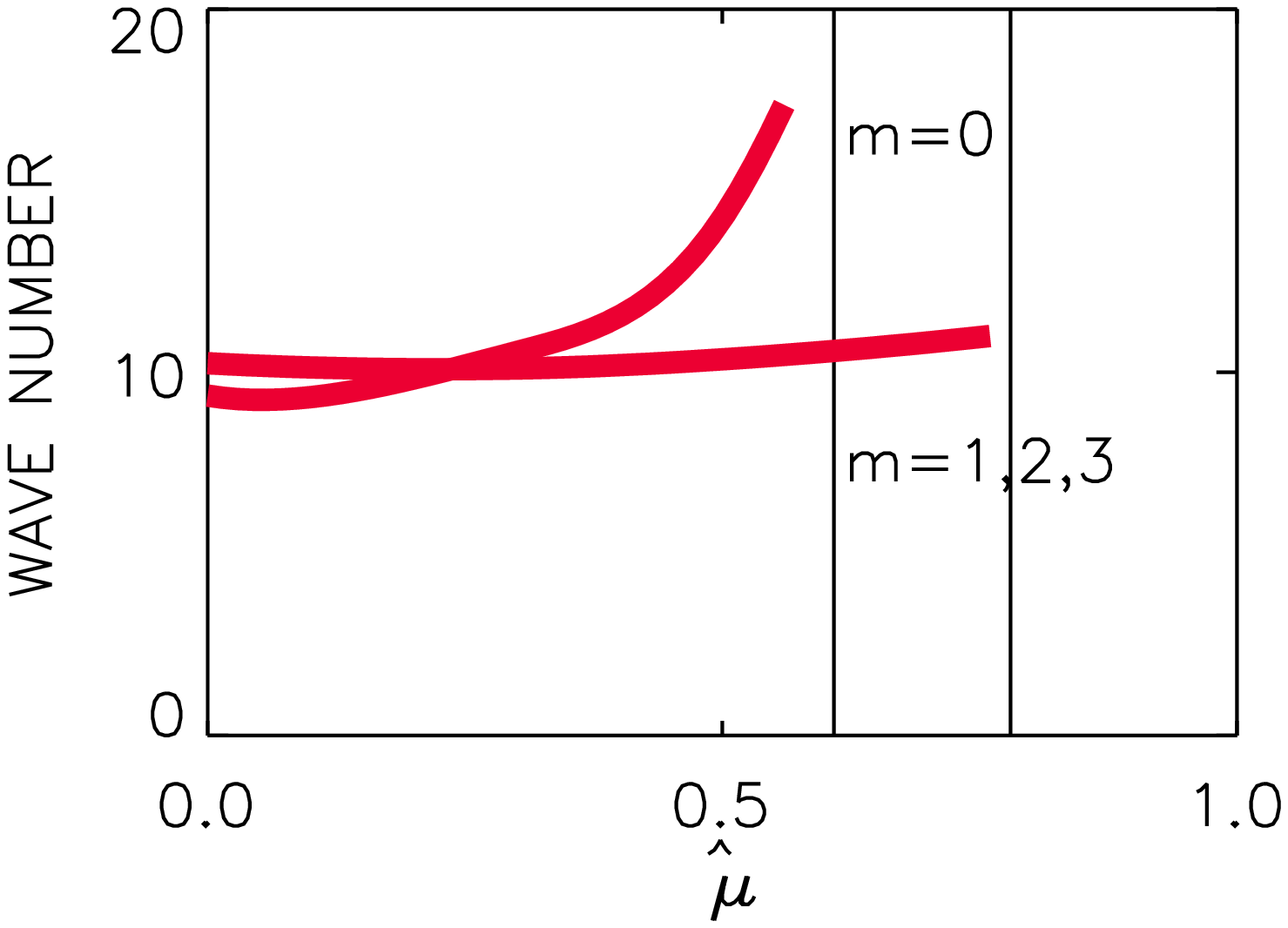}}
\caption{ Left: The marginal stability lines
 for $m\geq 0$. The vertical lines  
mark the Rayleigh line $\hat\mu=\hat \eta^2$ (left) and the outer limit line $\hat\mu=\hat \eta$ (right). Right: The vertical wave numbers are very large.
$\hat\eta=0.78$, $\rm Fr=0.5$. From Shalybkov \& R\"udiger (2005).}
\label{remu}
\end{figure}

The dependence of the critical Reynolds numbers
on $\hat\mu$ is given in
Fig.~\ref{remu} (left).
The already known marginal stability line  for $m=0$ and also the new 
lines  for $m>0$ are plotted. Note that the critical Reynolds numbers are remarkable small.
The axisymmetric disturbances are unstable only for $\hat\mu < \hat\eta^2$
in accordance to the Rayleigh condition.
The new nonaxisymmetric disturbances are also unstable in the 
interval $\hat\eta^2 < \hat\mu <\hat \eta $. The existence of the outer limit $\hat \mu=\hat \eta$ is a new and surprising finding. Even more flat rotation laws with $\hat \mu>\hat \eta$ are still stable (Shalybkov \& R\"udiger 2005). 

For $m>0$ the vertical wave number  only slightly depends on $m$
 (Fig.~\ref{remu}, right). The wave numbers are so large that the cells are flat in the axial direction. The vertical wave number does hardly depend on $\hat\mu$. 
\begin{figure}[htb]
\includegraphics[width=8cm,height=8.8cm]{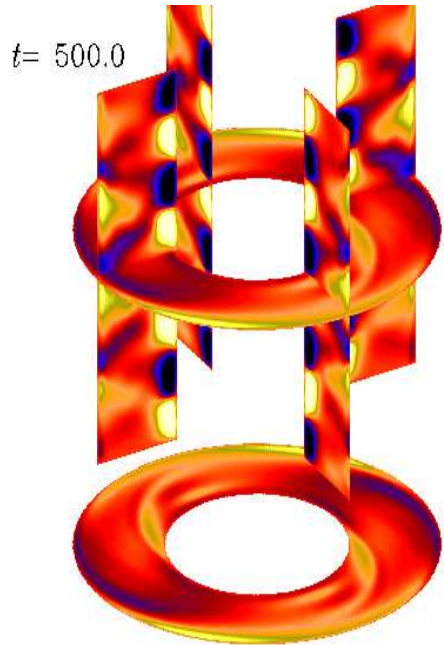}
\caption{The global stratorotational instability for   $\hat\mu=0.3$,   $\rm Fr=0.32$,  $\rm Re =500$ and $\rm Pr=10$. Note the $m=2$ azimuthal structure. From Brandenburg \& R\"udiger (2005).}
\label{axel}
\end{figure}

By Brandenburg \& R\"udiger (2005)  the global problem is treated  in Cartesian geometry with the nonlinear Pencil code. For a Reynolds number of about 500 a resolution of 128$^3$ mesh points was sufficient. Again $\hat \eta=0.5$. As expected, no instability
is found for $\hat\mu=0.6$. For smaller values ( $\hat\mu=0.2$) the flow is Rayleigh unstable. For  $\hat\mu=0.3$ the flow is indeed stratorotationally  unstable for nonaxisymmetric modes  also in the nonlinear regime. The modes are saturated and they are azimuthally drifting. Cross sections of $u_z$ are shown in Fig. \ref{axel} for $\hat\mu=0.3$ and $\rm Fr=0.32$. Note that there are 4 eddies in $z$-direction. The growth rate, however,   with $0.01 \Omega_{\rm in}$ proves to be rather small (see also Dubrulle et al. 2005).

\section{References}

\small
\noindent
Balbus, S.A.,  Terquem, C.,   {\it ApJ}, {\bf 552}, 235 (2002).\\ 
\noindent
Brandenburg, A., R\"udiger, {\it Astronomy \& Astrophysics}, in prep. (2005). \\
\noindent
Chandrasekhar, S., {\it Hydrodynamic \& Hydromagnetic Stability}, Oxford (1961).\\
\noindent
Donnelly, R.J., Ozima, M., {\it Phys. Rev. Lett.}, {\bf 4}, 497 (1960).\\
\noindent
Dubrulle, B., Marie, L., Normand, Ch. et al.,  {\it Astronomy \& Astrophysics}, {\bf 429}, 1 (2005). \\
\noindent
Hollerbach, R., R\"udiger, G., {\it Science}, in preparation (2005). \\ 
\noindent
Kurzweg, U.H., {\it J. Fluid Mech.}, {\bf 17}, 52 (1963).\\
\noindent
R\"udiger, G., Shalybkov, D., {\it Phys. Rev. E}, {\bf 66}, 016307 (2002).\\
\noindent
R\"udiger, G., Schultz, M., Shalybkov, D., {\it Phys. Rev. E}, {\bf 67}, 046312 (2003). \\
\noindent
R\"udiger, G., Shalybkov, D., {\it Phys. Rev. E}, {\bf 69}, 016303 (2004).\\ 
\noindent
R\"udiger, G., Hollerbach, R.,  {\it The magnetic universe}, 
Wiley (2004).\\
\noindent
Shalybkov, D., R\"udiger, G. Schultz, M.,  {\it Astronomy \& Astrophysics}, {\bf 395}, 339 (2005). \\
\noindent
Shalybkov, D., R\"udiger, G., {\it Astronomy \& Astrophysics}, subm. (2005). \\
\noindent
Velikhov, E.P., {\it Sov. Phys. JETP}, {\bf 9}, 995 (1959).\\
\noindent
Willis, A.P., Barenghi, C.F., {\it Astronomy \& Astrophysics}, {\bf 393}, 339 (2002).\\
\noindent
Withjack, E.M., Chen, C.F., {\it J. Fluid Mech.}, {\bf 66}, 725 (1974).\\ 
\noindent
Yavneh, I., McWilliams, J.C., Molemaker, M.J., {\it J. Fluid Mech.},
    {\bf 448}, 1 (2001).

\end{document}